\documentclass[12pt]{iopart}

\usepackage{graphics}

\newcommand{\bc}{\begin{center}}
\newcommand{\ec}{\end{center}}
\newcommand{\be}{\begin{equation}}
\newcommand{\ee}{\end{equation}}

\def\la{\mathrel{\mathchoice {\vcenter{\offinterlineskip\halign{\hfil
  $\displaystyle##$\hfil\cr<\cr\sim\cr}}}
  {\vcenter{\offinterlineskip\halign{\hfil$\textstyle##$\hfil\cr
  <\cr\sim\cr}}}
  {\vcenter{\offinterlineskip\halign{\hfil$\scriptstyle##$\hfil\cr
  <\cr\sim\cr}}}
  {\vcenter{\offinterlineskip\halign{\hfil$\scriptscriptstyle##$\hfil\cr
  <\cr\sim\cr}}}}}

\def\ga{\mathrel{\mathchoice {\vcenter{\offinterlineskip\halign{\hfil
  $\displaystyle##$\hfil\cr>\cr\sim\cr}}}
  {\vcenter{\offinterlineskip\halign{\hfil$\textstyle##$\hfil\cr
  >\cr\sim\cr}}}
  {\vcenter{\offinterlineskip\halign{\hfil$\scriptstyle##$\hfil\cr
  >\cr\sim\cr}}}
  {\vcenter{\offinterlineskip\halign{\hfil$\scriptscriptstyle##$\hfil\cr
  >\cr\sim\cr}}}}}

\begin{document}

\title[Light scattering by an elongated particle]
      {Light scattering by an elongated particle: spheroid versus
      infinite cylinder}

\author{ N V Voshchinnikov$\dag$ and V G Farafonov$\ddag$}

\address{$\dag$ Sobolev Astronomical Institute,
                 St.~Petersburg University,
                 Universitetskii pr. 28,
                 St.~Petersburg 198504, Russia}

\address{$\ddag$ St.~Petersburg University of Aerocosmic Instrumentation,
              Bol'shaya Morskaya ul. 67,
              St.~Petersburg 190000, Russia}

\eads{\mailto{nvv@astro.spbu.ru}, \mailto{far@softjoys.ru}}

\begin{abstract}
Using the method of separation of variables and a new approach to
calculations of the prolate spheroidal wave functions,
we study the optical properties of very elongated
(cigar-like)  spheroidal particles.

A comparison of extinction efficiency factors of
prolate spheroids and infinitely long circular cylinders
is made. For the normal and oblique incidence of radiation,
the efficiency factors for spheroids converge
to some limiting values with an increasing aspect ratio $a/b$ provided
particles of the same thickness are considered.
These values are close to, but do not coincide with the
factors for infinite cylinders.
The relative difference between factors for infinite cylinders
and elongated spheroids ($a/b \ga 5$) usually does not exceed
20\,\% if the following approximate relation between the
angle of incidence $\alpha~({\rm in~degrees})$ and
the particle refractive index $m=n+ki$ takes the place:
$\alpha  \ga 50 |m-1| + 5$ where
$1.2 \la n \la 2.0$ and $k \la 0.1$.

We show that the quasistatic approximation can be well used for
very elongated optically soft spheroids of large sizes.

\end{abstract}

\submitto{\MST}
\maketitle

\section{Introduction}

Rapid calculations of light scattering by non-spherical particles
are very important in many scientific and engineering applications
(see discussion in \cite{bh83}, \cite{mht00}).
The simplest model of non-spherical particles ---
an infinitely long circular cylinder is not physically reasonable.
However, it looks attractive to find cases when this model
could be useful
because the calculations for infinite cylinders are very simple.
Therefore, we compare the light scattering by
{\it elongated spheroids} and {\it infinite cylinders}.
Previously, such a comparison was made
by Martin \cite{mar78} and
Voshchinnikov \cite{v90b} for normal incidence of radiation
(perpendicular to the rotation axis of a particle,
$\alpha=90^)$).

Our consideration is based on the solution to the light scattering
problem for spheroidal particles by the
{\it Separation of Variables Method} (SVM)
(see \cite{f83}, \cite{vf93} for details).
A new type of expansions of the prolate wave functions
(Jaff\'e expansions  \cite{fv01})
opens a possibility to calculate the optical properties of
very elongated (cigar-like) particles.

In this paper, we study the optical properties of prolate
homogeneous spheroids with large aspect ratios and compare them with
those of infinite circular cylinders and of spheroids in quasistatic
approximation.

\section{Method}
\subsection{Prolate spheroid}

Prolate spheroid  is  obtained by an ellipse rotation around its major
axis. The shape of particle is characterized
by the  aspect ratio $a/b$ where $a$ and $b$ are the major and
minor semiaxes.
SVM's solutions to the electromagnetic problem for spheroids published
by Asano and Yamamoto~\cite{ay75} and Farafonov~\cite{f83}
(see also \cite{vf93})
are  fundamentally distinguished. In the former, the authors used
the Debye potentials to present the electromagnetic fields. This
approach is similar to the Mie solution for spheres.
Farafonov chose  special combinations of
the Debye and Hertz potentials, i.e. the potentials used
in the solutions for spheres and
infinitely long cylinders,  respectively. All the electromagnetic fields were divided
into two parts: the axisymmetric part not depending on the azimuthal
angle $\varphi$, and the non-axisymmetric part when the integration
over $\varphi$ gives zero.

The computational efficiency of the Asano and Yamamoto's
and Farafonov's solutions can be compared in the following way
(see \cite{vf93} for details).  Let $N$ is the number of terms in sums
which give the efficiency factors for extinction and scattering
$Q_{\rm ext}, \, Q_{\rm sca}$ with some accuracy
in the Farafonov's solution. Then, to obtain the results with the same accuracy,
the solution of Asano and Yamamoto requires $\approx 2N$ and $\approx 5N$
terms if the  aspect ratio
$a/b \approx 2$ and $a/b \approx 10$, correspondingly. Because the
computational time is proportional to $N^3$ in this case
and $t \propto N^2$ for the Farafonov's solution,
the advantage of the latter is evident, especially for large aspect ratios.

It should also be noted that the convergence of  Farafonov's solution
for spheroids follows that of the Mie  solution for spheres
(see Table~2 in  Voshchinnikov, \cite{v96}).

It must be emphasized that the previous calculations for spheroidal particles with
the Asano and Yamamoto's solution were restricted by the aspect
ratios $a/b \leq 5$ (size parameter $2 \pi a/\lambda \leq 30$) \cite{a79}
or $a/b = 10$ ($2 \pi a/\lambda \la 10$) \cite{ks93}.

The particle size can also be specified by the parameters
$x_V = 2 \pi r_V/ \lambda$ ($r_V$ is the radius of
a sphere of volume equal to that of spheroid, $\lambda$ the wavelength
of incident radiation) or
$c =  2 \pi/\lambda \cdot d/2$ ($d$ is the focal distance of a spheroid).
The expressions relating different parameters have the form
\be
\frac{2\pi a}{\lambda} =  c \xi_0
\ee
\be
x_{V} = \frac{2\pi r_{V}}{\lambda} =
\frac{2\pi a}{\lambda} \left ( \frac{a}{b} \right )^{-2/3}
\label{x_v}\ee
where $r_V^3  = a b^2$.
The parameter $\xi_0$ depends only on the aspect ratio $a/b$
\be
\xi_0 = \left (\frac{a}{b} \right )
\left [ \left ( \frac{a}{b} \right )^{2} - 1 \right ]^{-1/2}.
\ee

All cross-sections $C$ (extinction, scattering etc)
are connected with the corresponding
efficiency factors $Q$ via the relation
\be
C = G Q \label{c}
\ee
where $G$ is the ``viewing'' geometrical cross-section of a particle
(the area of the particle shadow).
The factors $Q$ and geometrical cross-sections $G$ depend on
the angle of incidence $\alpha$ (the angle
between the direction of radiation incidence and the particle
rotation axis, $0^0 \leq \alpha \leq 90^0$).
The  geometrical cross-section of prolate spheroid is
\begin{equation}
G(\alpha) = \pi b \left(a^2\sin^2\alpha
            + b^2\cos^2\alpha\right)^{1/2}.
\label{g_p}\end{equation}

In the case of oblique radiation incidence ($\alpha \neq 0^0$),
the factors depend on the state of polarization of incident radiation:
$Q^{\rm TM, \, TE}$.
The superscript TM (TE) is related to the case
when the electric vector $\vec{E}$ of the incident radiation
is parallel (perpendicular) to the plane defined
by the rotation axis of a spheroid and the wave vector.
Expressions for the factors  can be found in \cite{vf93}.

In order to compare the optical properties of the particles of different
shapes it is convenient to consider the ratios of the cross-sections
for spheroids to the geometrical cross-sections of the
equal volume spheres, $C/\pi r^{2}_{V}$.
 They can be found as
\be
\frac{C}{\pi r^{2}_{V}} = \frac{[(a/b)^2\sin^2\alpha
                             +\cos^2\alpha]^{1/2}}{(a/b)^{2/3}} \, Q.
\label{crv_pro}
\ee

\subsection{Infinite cylinder}

The extinction, scattering and absorption cross-sections of an infinite cylinder
are, of course, infinite.
In order to keep some physical sense the infinite cylinder is replaced by
a very-very long finite cylinder and the cross-sections are
calculated {\it per unit length} of an infinite cylinder
(see \cite{bh83}, p.~203).

For the infinite  circular cylinder, the efficiencies are usually defined
as the corresponding cross-sections per unit length, $c_{\rm cyl}$,
divided by $2 a_{\rm cyl}$, the diameter of the cylinder \cite{lindg66}.

For a finite cylinder of length $2L$ and radius $a_{\rm cyl}$, $Q$ is
the ratio of the cross-section $C_{\rm cyl}$ to the normally projected
geometric area of the cylinder, $2a_{\rm cyl}2L$.

The extinction efficiencies are considered for two cases of polarization
of incident radiation: E case (TM mode) and H case (TE mode) \cite{bh83}, \cite{lindg66}
\be
Q^E_{\rm ext} = \frac{c^E_{\rm cyl}}{2a_{\rm cyl}} =
 \frac{C^E_{\rm cyl}}{2a_{\rm cyl} 2L}
= \frac{2}{x_{\rm cyl}} {\rm Re} \left\{b^E_0 + 2 \sum^\infty_{n=1}
b^E_n \right\}
\ee
\be
Q^H_{\rm ext} = \frac{c^H_{\rm cyl}}{2a_{\rm cyl}} =
 \frac{C^H_{\rm cyl}}{2a_{\rm cyl} 2L}
= \frac{2}{x_{\rm cyl}} {\rm Re} \left\{a^H_0 + 2 \sum^\infty_{n=1}
a^H_n \right\}
\ee
where $x_{\rm cyl} = 2 \pi a_{\rm cyl} / \lambda$.
The coefficients $a^H_n$ and $b^E_n$ depend on the complex refractive index $m=n+ki$,
size parameter $x_{\rm cyl}$ and the angle $\alpha$ ($0^0 < \alpha \leq 90^0$).
The expressions for coefficients are given in  \cite{lindg66}.

Because of different definitions of geometrical cross-sections for spheroids
and cylinders, the comparison of factors requires the normalization
\be
Q^{\rm TM, \, TE}_{\rm sph} = \frac{Q^{\rm E, \, H}_{\rm cyl}}{\sin \alpha}
\ee
where $\alpha \neq 0^0$.

\section{Numerical results}

In previous attempts to find the limits of applications of the  model
of infinite cylinders (\cite{mar78}, \cite{v90b}), the particles of the
same thickness were considered, i.e. spheroids and
cylinders had the equal parameters
$2 \pi b/\lambda = x_{\rm cyl} =  2 \pi a_{\rm cyl}/\lambda$.
Martin~\cite{mar78} notes that the extinction factors
for spheroids become to resemble those for cylinders if $a/b \ga 4$
for the normal incidence of radiation ($\alpha=90^0$).
But in order to align the peaks in extinction, the $x$ scale for cylinders
was stretched by a factor 1.13.

We suggest another way for comparison
of the efficiency factors: to put equal
the volume and the aspect ratio of a spheroid
and a very long cylinder, respectively, i.e.
\be
V_{\rm sph} = V_{\rm cyl}, \,\,\,\,\,\,\,\,\,\,\,\,\,\,\,\,\,\,\,\,\,
\frac{a}{b} = \frac{L}{a_{\rm cyl}}. \label{cond}
\ee
From equation~(\ref{cond}), we obtain the relation between
the size parameters for infinite  cylinder  and prolate spheroid
\be
\frac{2\pi a_{\rm cyl}}{\lambda} =
\frac{2\pi b}{\lambda} \, \left ( \frac{2}{3} \right )^{1/3}.
\label{rel} \ee
Note that the scaling factor arising in equation~(\ref{rel})
[$(3/2)^{1/3} \approx 1.145$] is close to that found
empirically in \cite{mar78}.

\subsection{Normal and oblique incidence of radiation}

\begin{figure}[htb]
\bc
\resizebox{\hsize}{!}{\includegraphics{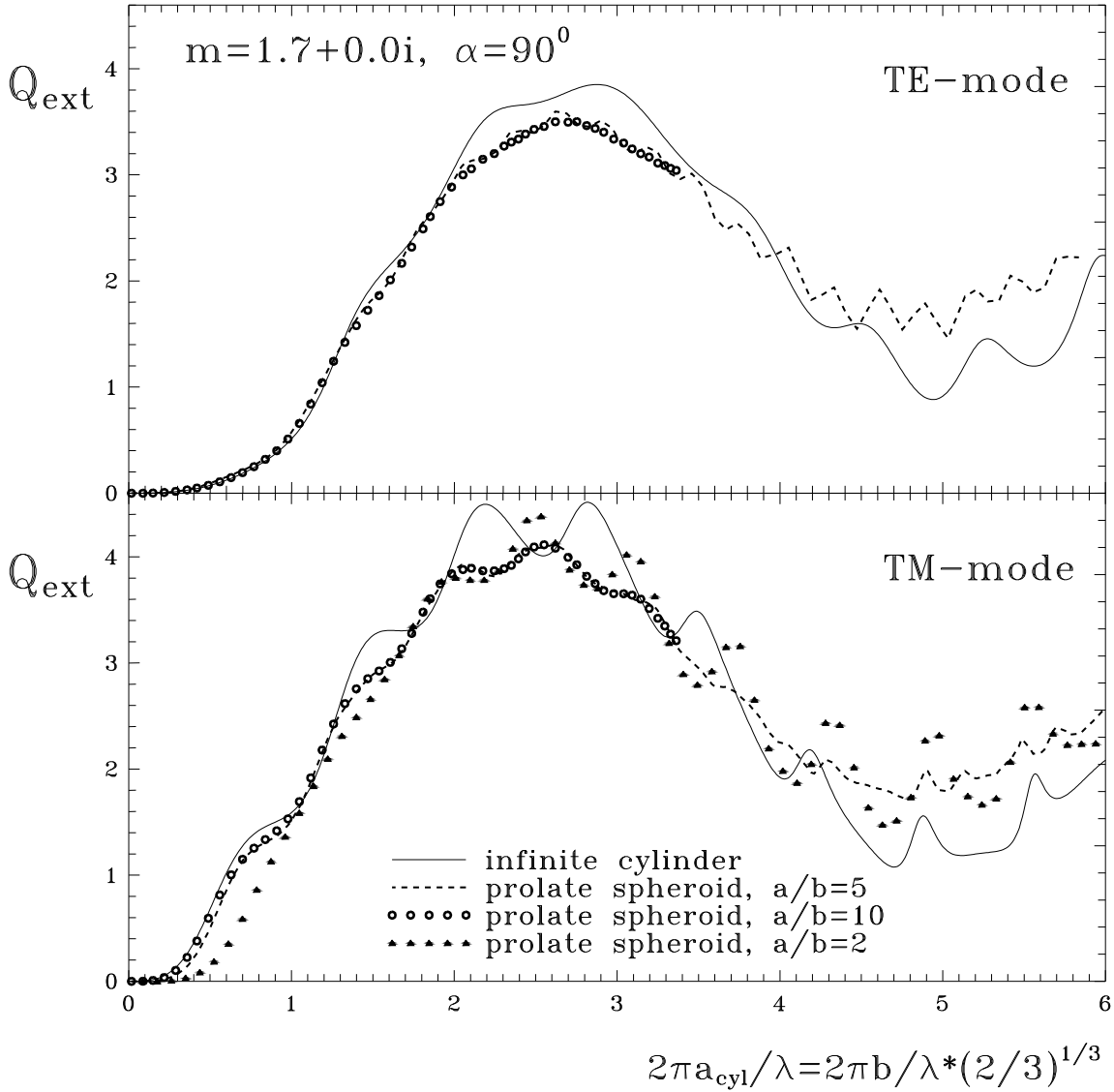}}
\caption
{
Extinction efficiency factors (TM and TE modes)
for prolate spheroids ($Q_{\rm ext} = C_{\rm ext}/G(\alpha)$)
and infinite cylinders
($Q_{\rm ext} = C_{\rm ext}/(2 a_{\rm cyl} \sin \alpha)$)
as a function of the size parameter
$2 \pi a_{\rm cyl}/\lambda = 2 \pi b/\lambda \cdot (2/3)^{1/3}$.
}
\label{f1}
\ec
\end{figure}

\begin{figure}
\bc
\resizebox{15.1cm}{!}{\includegraphics{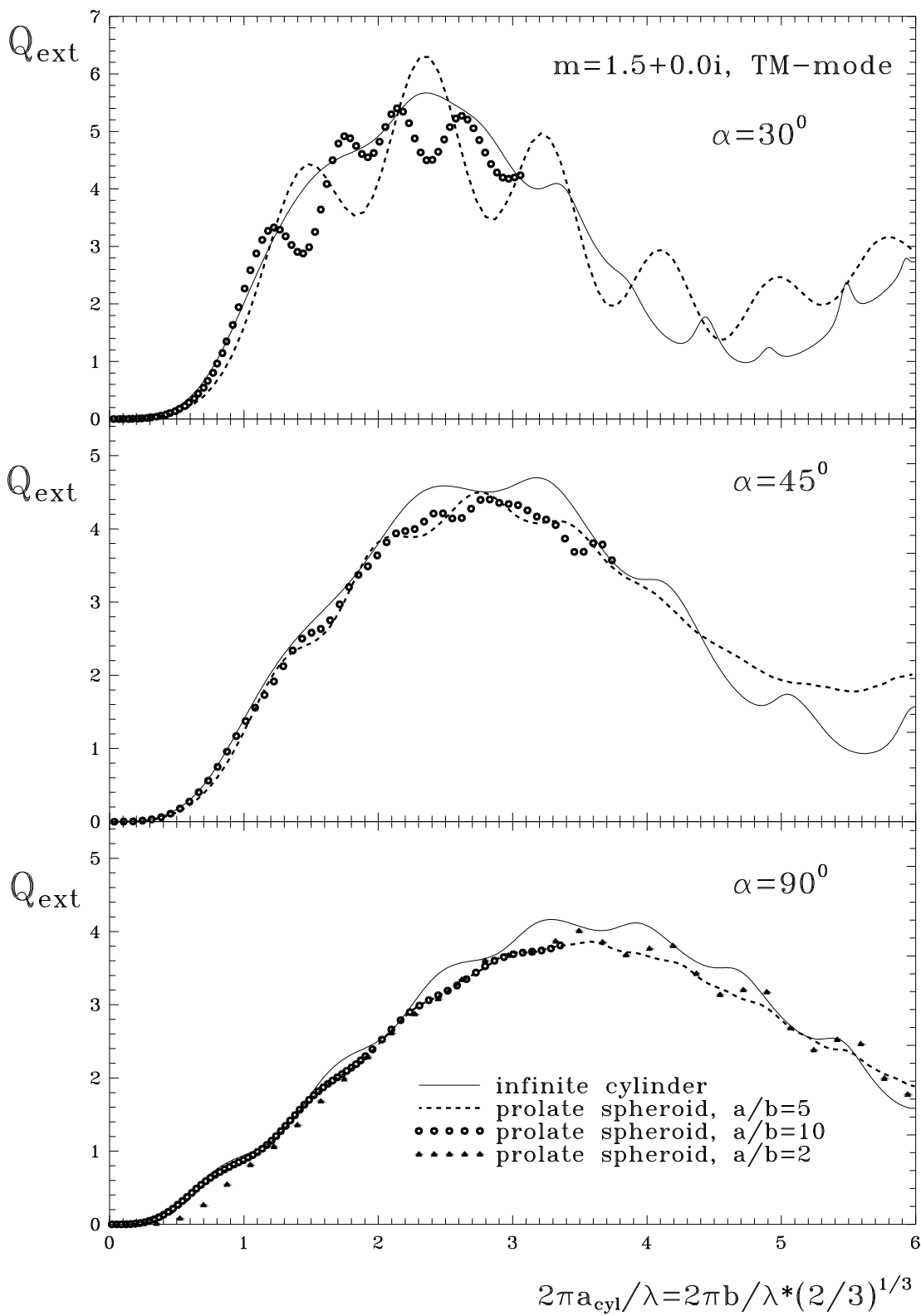}}
\caption
{
Extinction efficiency factors (TM mode)
for prolate spheroids ($Q_{\rm ext} = C_{\rm ext}/G(\alpha)$)
and infinite cylinders
($Q_{\rm ext} = C_{\rm ext}/(2 a_{\rm cyl} \sin \alpha)$)
as a function of the size parameter
$2 \pi a_{\rm cyl}/\lambda = 2 \pi b/\lambda \cdot (2/3)^{1/3}$.
}
\label{f2}
\ec
\end{figure}

The extinction efficiency factors for spheroids and infinite cylinders
are plotted in figure~\ref{f1} for the case of  normal incidence of radiation
(perpendicular to the rotation axis, $\alpha =90^\circ$).
There is a {\it similarity} of the behaviour
of the efficiency factors: both TM and TE modes converge to some
limiting values which are close to those of infinite cylinders
but do not coincide with them.
This occurs if the size parameter is defined by the relation~(\ref{rel})
and the aspect ratio $a/b \ga 5$ (a noticeable difference in factors
for $a/b=2$ it is seen in lower panels of Figs.~\ref{f1}, \ref{f2}).

If we reduce the angle $\alpha$ (figure~\ref{f2}, upper and middle panels),
the similarity in behaviour of factors remains although the ripple-like
structure changes. The likeness disappears if we approach the grazing case:
the first peak for cylinders occurs at smaller size parameters than
for spheroids (see also figure~\ref{f5}). For refractive index $m=1.5+0.0i$,
it takes place if the angle $\alpha$ becomes smaller than $\sim 30^0$.
Because the ripples reduce with the decrease of the
real part of refractive index $n$
or the increase of its imaginary part $k$, in these cases the
infinite cylinders can be used instead of elongated spheroids at smaller
values of $\alpha$. However, the difference becomes very large for
$\alpha \la 15^0$. If $1.2 \la n \la 2.0$ and $k \la 0.1$,
the linear relation between the
angle of incidence $\alpha~({\rm in~degrees})$ and
the particle refractive index $m=n+ki$ takes the place:
$\alpha  \ga 50 |m-1| + 5$. Then
the relative discrepancy between factors for infinite cylinders
and elongated spheroids ($a/b \ga 5$)  does not exceed
20\,\% near the  first maximum.

Anyway, we can expect the resemblance of the extinction properties
of very elongated particles (spheroids and cylinders) in the case
of radiation incidence close to normal if the size parameters
are recalculated according to equation~(\ref{rel}).

\subsection{Parallel incidence of radiation}

\begin{figure}[htb]
\bc
\resizebox{\hsize}{!}{\includegraphics{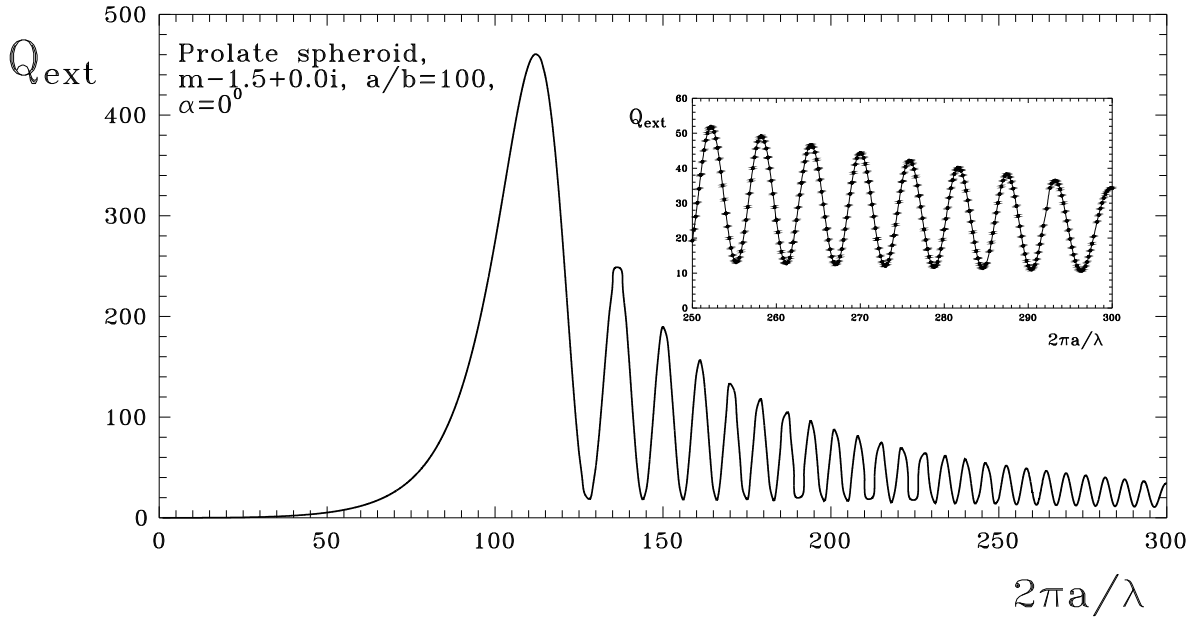}}
\caption
{Extinction efficiency factors
for elongated spheroidal particles in dependence of size parameter
$2\pi a / \lambda$.
}
\label{f3}
\ec
\end{figure}

\begin{figure}
\bc
\resizebox{\hsize}{!}{\includegraphics{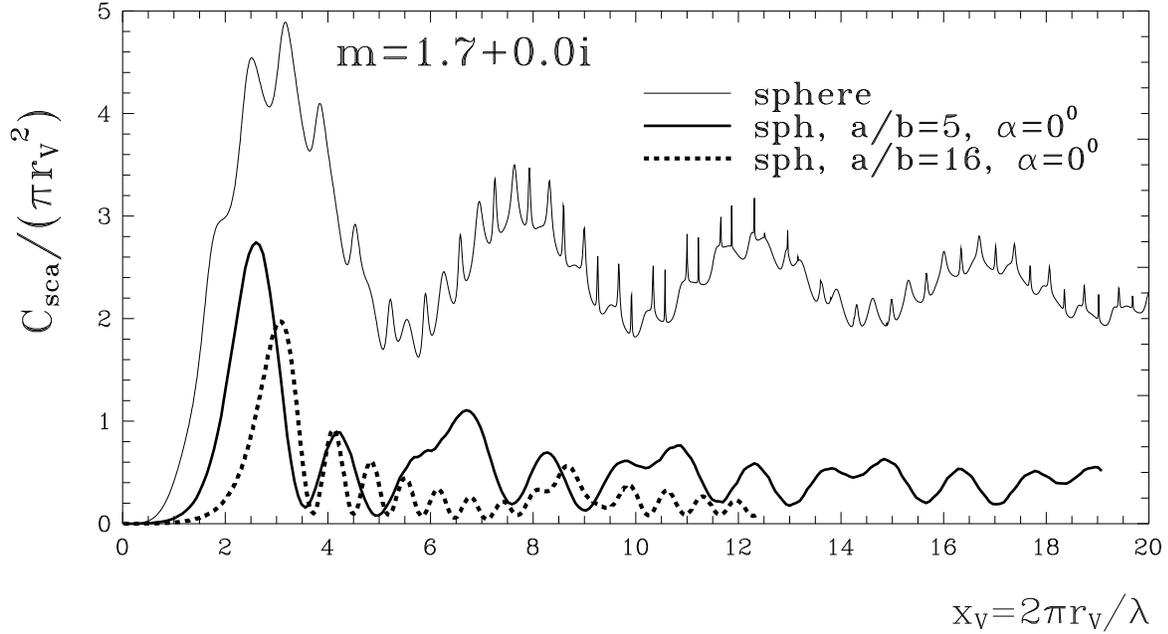}}
\caption{
Normalized extinction  cross-sections
in dependence of the size  parameter $2\pi r_V /\lambda$
for spheres and prolate spheroids.
}
\label{f4}
\ec
\end{figure}

Figure~\ref{f3} shows the extinction efficiency factors for very long spheroids
in the case of the incidence of radiation along the rotation axis of the
particles. The behaviour of the factors is rather regular, and their values
smoothly reduce with the size parameter.
There are 25 maxima on the interval $2\pi a/\lambda= 0-300$ that
is totally distinct from the extinction by
spherical particles (cf. figure~\ref{f4}).
The positions of the maxima for large particles are determined
by the path of light inside the spheroids (the phase shift of the central ray).
Note that the ``equivolume'' size of particles $x_V$ considered on
figure~\ref{f3} is moderate: from equation~(\ref{x_v}) follows that
$x_V \approx 13.92$ if $2\pi a/\lambda= 300$ and $a/b=100$.
But the path of rays inside spheroidal particle is in
$300/13.92 \approx 21.5$ times longer in comparison with spherical particle.

The values of factors on figure~\ref{f3} are very large.
This is the result of normalization by
geometrical cross-section which is small in this case and  is equal to
$G(0^0) = \pi b^2$ (see equation~(\ref{g_p})).
Indicate that the limiting value of extinction factors for particles
of any shape must be equal to 2 (``extinction paradox'')
but as it is seen from the insertion on figure~\ref{f3},
this condition is far to be satisfied
although the tendency for reduction of $Q_{\rm ext}$ is observed.

Figure~\ref{f4} shows the normalized extinction cross-sections for
spheres and prolate spheroids at parallel incidence of radiation
($\alpha=0^0$). As follows from figure~\ref{f4}  spheres of the same volume
scatter more radiation than spheroids but the situation changes
in the case of another orientation of spheroids.
Note also on the absence of ripple-like structure on curves plotted for spheroids.

\begin{figure}
\bc
\resizebox{\hsize}{!}{\includegraphics{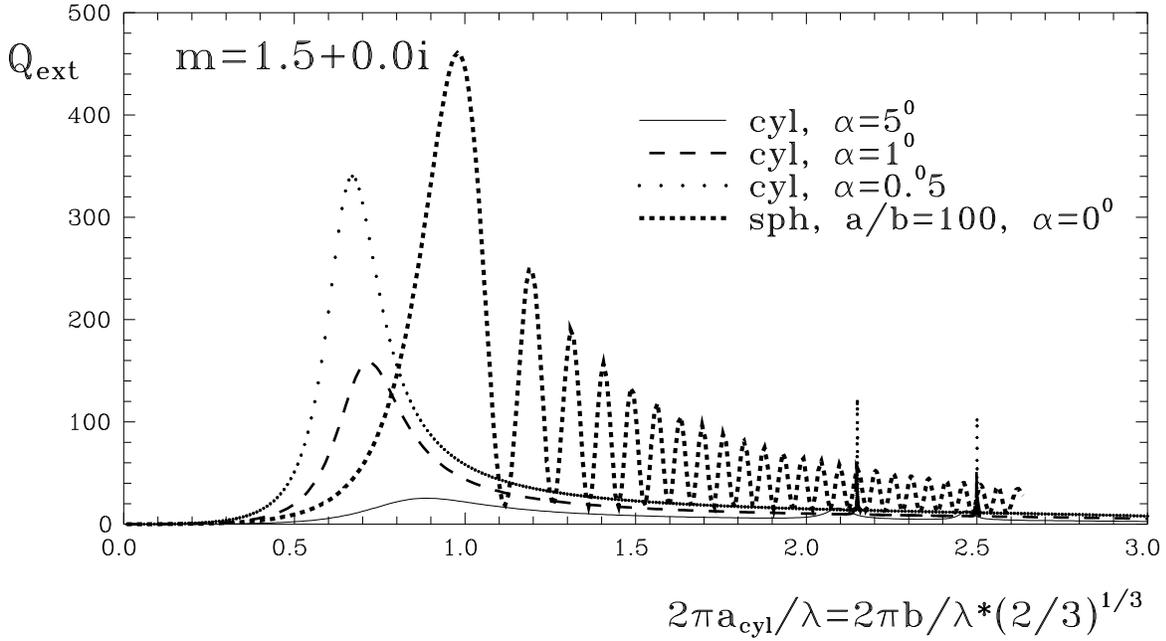}}
\caption{
Extinction efficiency factors (TM mode)
for prolate spheroids ($Q_{\rm ext} = C_{\rm ext}/G(\alpha)$)
and infinite cylinders
($Q_{\rm ext} = C_{\rm ext}/(2 a_{\rm cyl} \sin \alpha)$)
as a function of the size parameter
$2 \pi a_{\rm cyl}/\lambda = 2 \pi b/\lambda \cdot (2/3)^{1/3}$.
}
\label{f5}
\ec
\end{figure}

The ripple-like structure disappear on the extinction curves
for cylinders as well in the case
of nearly grazing incidence of radiation (figure~\ref{f5}).
From figure~\ref{f5},  it is seen that the similarity between elongated
spheroids and infinite cylinders is absent.
For tangentially incident radiation, the positions and strengths of maxima for
infinite cylinders do not coincide with those for spheroids at $\alpha =0^0$.

\subsection{Quasistatic approximation}

The optical properties of extremely prolate and oblate
particles may be approximately calculated using
the {\it quasistatic approximation}. This is a generalization
of the Rayleigh and Rayleigh-Gans approximations
when the electromagnetic field inside a particle
is replaced by the incident radiation field
(as in the Rayleigh-Gans approximation), taking into account
the polarizability of the particle (as in the Rayleigh approximation).
The expressions for the efficiency factors and amplitude matrices
were obtained in \cite{far94b}.
The range of validity of the quasistatic approximation is
discussed in \cite{vf00}.

\begin{figure}
\bc
\resizebox{\hsize}{!}{\includegraphics{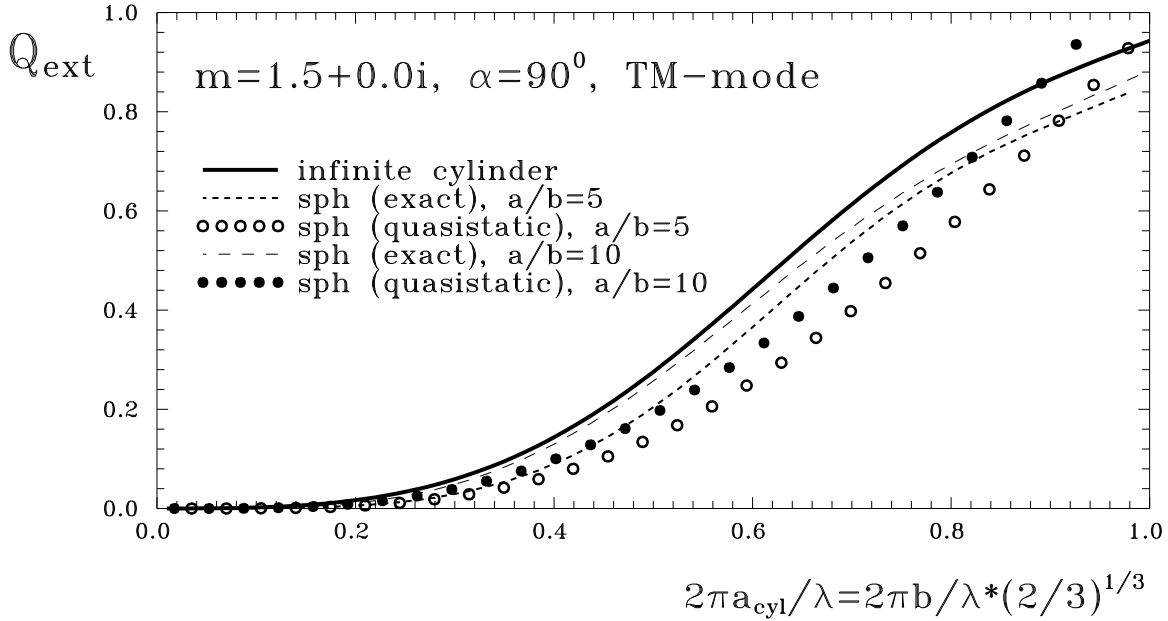}}
\caption{
    The extinction efficiencies
     for prolate spheroids and infinite cylinders.
    The results for spheroids were calculated using exact
    solution and quasistatic approximation.
}
\label{f6}
\ec
\end{figure}

\begin{figure}
\bc
\resizebox{\hsize}{!}{\includegraphics{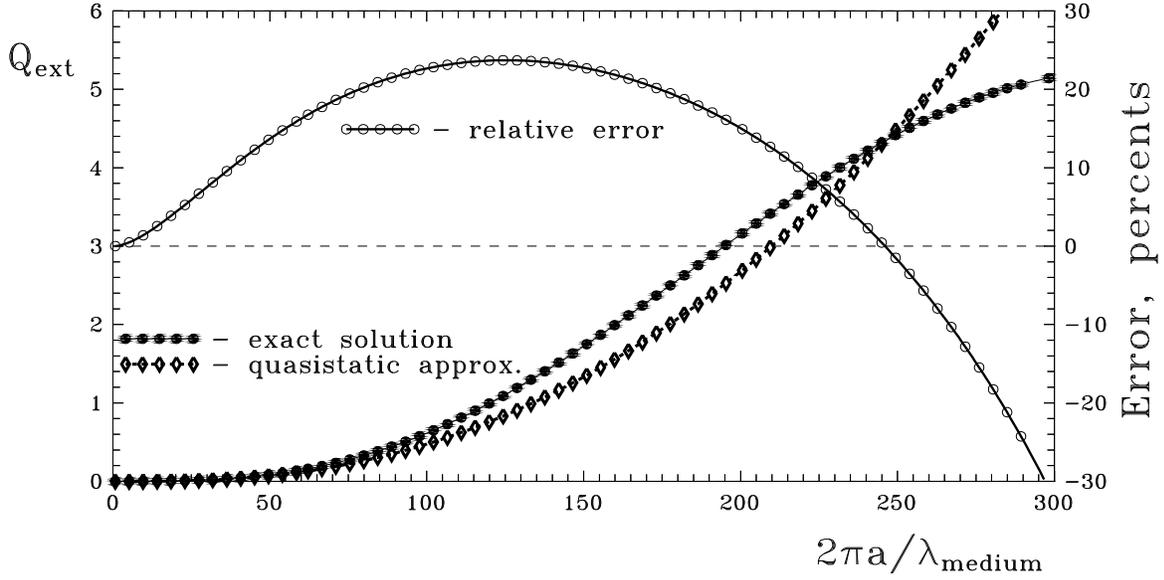}}
\caption{
    The extinction efficiency factors (left $y$-axis)
    for prolate spheroids ($m_{\rm medium} = 1.48+0.0i$,
    $m_{\rm cavity} = 1.49+0.0i$, $a/b=10$,
     $\alpha=0^0$) in dependence of the size  parameter $2\pi a /\lambda_{\rm medium}$.
    The calculations were made  using exact
    solution and quasistatic approximation.
    The relative  error of the calculations of factors in
    quasistatic approximation is given by the right $y$-axis.
}
\label{f7}
\ec
\end{figure}

Figures~\ref{f6} and \ref{f7} show the extinction efficiency factors
for spheroids calculated with the exact solution and the quasistatic
approximation in the case of normal ($\alpha=90^0$) and parallel
($\alpha=0^0$) incident radiation. As we consider non-absorbing particles,
the extinction efficiencies coincide with the scattering ones which
are in the quasistatic approximation:
\be
\fl
Q^{\rm TM}_{\rm sca}(0^0) =
\frac{c^4 \xi_0^2 (\xi_0^2-1)}{9 \pi }
      |\tilde{\alpha}_1|^2
      \int_0^{2 \pi} \int_0^{\pi}
      (\sin^2 \varphi + \cos^2 \theta \cos^2 \varphi) \,
      G^2(u) \sin \theta {\rm d}\theta {\rm d}\varphi
\ee
\be
Q^{\rm TM}_{\rm sca}(90^0) =
\frac{c^4 \xi_0^2 (\xi_0^2-1)}{9 \pi }
      |\tilde{\alpha}_3|^2
      \int_0^{2 \pi} \int_0^{\pi}
      G^2(u) \sin^3 \theta {\rm d}\theta {\rm d}\varphi
\ee
where $\tilde{\alpha}_1$ and $\tilde{\alpha}_3$ are the polarizabilities,
\be G(u) = \frac{3}{u^3}(\sin u - u \cos u) \ee
and
\be u = c \xi_0 \sqrt{(\cos \theta - 1)^2 +
\left ( \frac{a}{b} \right )^{-2}\sin^2 \theta} \ee
if $\alpha=0^0$ and
\be u = c \xi_0 \sqrt{\cos^2 \theta +
\left ( \frac{a}{b} \right )^{-2}
(\sin^2 \theta + 1 - 2 \sin \theta \cos \varphi)} \ee
if $\alpha=90^0$.

Figure~\ref{f6} shows the enlarged part of figure~\ref{f2} (lower panel)
for small size parameters. Here,
the results for quasistatic approximation are also plotted.
It is seen that this approximation describes rather well
the behaviour of efficiencies for relatively large size parameters
(the value $2 \pi a_{\rm cyl} / \lambda=1$ corresponds to
$2 \pi a / \lambda=5.725$ for $a/b=5$ and
$2 \pi a / \lambda=11.45$ for $a/b=10$).

The extinction efficiencies for optically soft particles are drawn in
figure~\ref{f7}.
The results given here show the light scattering
by a spheroidal cavity ($m_{\rm cavity} = 1.49+0.0i$) in glass
($m_{\rm medium} = 1.48+0.0i$). As follows from figure~\ref{f7},
the quasistatic approximation allows to calculate the optical properties
with the relative error smaller than $\sim 25\,\%$ in a wide range
of particle sizes.

The range of validity of the quasistatic approximation is
discussed in \cite{vf00}.
It was  found that a maximum parameter  $x_V$ for which
the quasistatic approximation and exact theory
yielded the results coinciding within
1\%, can be described by the following approximate formulae:
\be
x_V \la \frac{0.02 \ln (a/b) + 0.13} {(n - 1)^{0.30}}
\ee
for prolate spheroids and $\alpha =0^{\circ}$,
\be
x_V \la \frac{0.10}{(n - 1)^{0.13 \ln (a/b) + 0.29}}
\ee
for prolate spheroids and $\alpha =90^{\circ}$,
\be
x_V \la \frac{0.11}{(n - 1)^{0.09 \ln (a/b) + 0.26}}
\ee
for oblate spheroids and $\alpha =0^{\circ}$,
\be
x_V \la \frac{0.06 \ln (a/b) + 0.12} {(n - 1)^{0.23}}
\ee
for oblate spheroids and $\alpha =90^{\circ}$.

\section{Conclusions}

We applied a new method of calculations of the spheroidal wave functions
to the study of the optical properties of very elongated
(cigar-like)  spheroids. New approach allowed to
compare light scattering by prolate spheroids and infinitely long circular cylinders
and investigate the applicability of  the quasistatic approximation
for very elongated spheroids.

It is found that the  efficiency factors for spheroids and cylinders
have quite similar behaviour for the normal and oblique
incidence of radiation,
if the aspect ratio of spheroids $a/b \ga 5$.
The resemblance of factors arises
provided spheroids and very long cylinders of the same volume and
aspect ratio are considered.
The following approximate relation between the
angle of incidence $\alpha~({\rm in~degrees})$ and
the particle refractive index $m=n+ki$ takes the place:
$\alpha  \ga 50 |m-1| + 5$ where
$1.2 \la n \la 2.0$ and $k \la 0.1$.
In this case,
the relative discrepancy between factors for infinite cylinders
and elongated spheroids ($a/b \ga 5$)  does not exceed
20\,\% near the  first maximum.

It is shown that the quasistatic approximation rather well describes
the extinction by very elongated optically soft spheroids
of large sizes.

\ack{The authors are thankful to Vladimir Il'in for useful
comments. The work was partly supported by the INTAS grant 99/652,
grant for scientific school on theoretical astrophysics,
and the program ``Astronomy'' of the Russian Federal Government.}

\end{document}